\documentclass[]{spie}  
\usepackage[]{graphicx}
\usepackage[breaklinks, colorlinks, citecolor=blue, linkcolor=MyBlue, urlcolor=RoyalPurple, colorlinks=true, linkcolor=blue, debug, baseurl=' ']{hyperref}

\title{The NIKA2 commissioning campaign: performance and first results} 

\author{
A.~Catalano\supit{a},
R.~Adam\supit{b,}\supit{a},
P.~Ade\supit{e},
P.~Andr\'e\supit{d},
H.~Aussel\supit{d},
A.~Beelen\supit{f},
A.~Beno\^it\supit{g},
A.~Bideaud\supit{g},
N.~Billot\supit{g},
O.~Bourrion\supit{a},
M.~Calvo\supit{g},
G.~Coiffard\supit{c},
B.~Comis\supit{a},
F.-X.~D\'esert\supit{l},
S.~Doyle\supit{e},
J.~Goupy\supit{g},
C.~Kramer\supit{h},
G.~Lagache\supit{r},
S.~Leclercq\supit{c},
J.F.~Lestrade\supit{q},
J.F.~Mac\'ias-P\'erez\supit{a},
A.~Maury\supit{d},
P.~Mauskopf\supit{e,}\supit{n},
F.~Mayet\supit{a},
A.~Monfardini\supit{g},
F.~Pajot\supit{f},
E.~Pascale\supit{e},
L.~Perotto\supit{a},
G.~Pisano\supit{e},
N.~Ponthieu\supit{l},
V.~Rev\'eret\supit{d},
A.~Ritacco\supit{a},
L.~Rodriguez\supit{d},
C.~Romero\supit{c},
H.~Roussel\supit{s}, 
F.~Ruppin\supit{a},
K.F.~Schuster\supit{c},
A.~Sievers\supit{h},
J.~Soler\supit{f},  
S.~Triqueneaux\supit{g},
C.~Tucker\supit{e},
R.~Zylka\supit{c},
\skiplinehalf
\small
\supit{a}Laboratoire de Physique Subatomique et de Cosmologie, Universit\'e Grenoble Alpes, CNRS/IN2P3, 53, rue des Martyrs, Grenoble, France \\
\supit{b}Laboratoire Lagrange, Universit\'e C\^ote d'Azur, Observatoire de la C\^ote d'Azur, CNRS, Blvd de l'Observatoire, CS 34229, 06304 Nice cedex 4, France \\
\supit{c}Institut de RadioAstronomie Millim\'etrique (IRAM), Grenoble, France \\
\supit{d}Laboratoire AIM, CEA/IRFU, CNRS/INSU, Université Paris Diderot, CEA-Saclay, 91191 Gif-Sur-Yvette, France \\
\supit{e}Astronomy Instrumentation Group, University of Cardiff, UK \\
\supit{f}Institut d'Astrophysique Spatiale (IAS), CNRS and Universit\'e Paris Sud, Orsay, France \\
\supit{g}Institut N\'eel, CNRS and Universit\'e Grenoble Alpes, France \\
\supit{h}Institut de RadioAstronomie Millim\'etrique (IRAM), Granada, Spain \\
\supit{i}Dipartimento di Fisica, Sapienza Universit\'a di Roma, Piazzale Aldo Moro 5, I-00185 Roma, Italy \\
\supit{l} Institut de Plan\'etologie et d'Astrophysique de Grenoble (IPAG), CNRS, Universit\'e Grenoble Alpes, F-38000 Grenoble, France  \\
\supit{m}Aix Marseille Universit\'e, CNRS, LAM (Laboratoire d'Astrophysique de Marseille) UMR 7326, 13388, Marseille, France \\
\supit{n}School of Earth and Space Exploration and Department of Physics, Arizona State University, Tempe, AZ 85287 \\
\supit{o}Universit\'e de Toulouse, UPS-OMP, Institut de Recherche en Astrophysique et Plan\'etologie (IRAP), Toulouse, France \\
\supit{p}CNRS, IRAP, 9 Av. colonel Roche, BP 44346, F-31028 Toulouse cedex 4, France \\
\supit{q} LERMA, CNRS, Observatoire de Paris, 61 Avenue de l'Observatoire, Paris, France \\
\supit{r} Aix Marseille Universit\'e, CNRS, LAM (Laboratoire d'Astrophysique de Marseille) UMR 7326, 13388, Marseille, France  \\
\supit{s} Institut d'Astrophysique de Paris, CNRS (UMR7095), 98 bis Boulevard Arago, F-75014, Paris, France \\
}

\authorinfo{Further author information: (Send correspondence to Andrea Catalano)\\Andrea Catalano: E-mail: catalano@lpsc.in2p3.fr, Telephone: +33-476284152}
 \normalsize

\begin{document} 
  \maketitle 
\begin{abstract}
The New IRAM KID Array 2 (NIKA~2) is a dual-band camera operating with three frequency-multiplexed kilopixels arrays of Lumped Element Kinetic Inductance Detectors (LEKID) cooled at 150~mK. NIKA~2 is designed to observe the intensity and polarisation of the sky at 1.15 and 2.0~mm wavelength from the IRAM 30~m telescope. The NIKA 2 instrument represents a huge step in performance as compared to the NIKA pathfinder instrument, which has already shown state-of-the-art detector and photometric performance. After the commissioning planned to be accomplished at the end of  2016, NIKA~2 will be an IRAM resident instrument for the next ten years or more. NIKA~2 should allow the astrophysical community to tackle a large number of open questions reaching from the role of  the Galactic magnetic field in star formation to the discrepancy between cluster-based and CMB-based cosmology possibly induced by the unknown cluster physics. We present an overview of the commissioning phase together with some first results. 
\end{abstract}



\begin{table}[h!]
\label{table:sch}
\begin{center}   
\begin{tabular}{|c|} 
\hline
\hline
{\bf NIKA project schedule} \\
\hline
2009-2015 NIKA (opened to community in 2014).  \\
2012-2015 NIKA2 development. \\
February 2015 new telescope cabin optics. \\
October 2015 installation of NIKA2. \\
2016 commissioning and upgrades. \\
2017 opening to community.  \\
\hline 
\hline
\end{tabular}
\end{center}
\caption{Schedule of the NIKA project: NIKA is a collaboration between several laboratories. Institut N\'eel, LPSC, IPAG,  and IRAM are the main contributors.} 
\end{table} 

\begin{table}[t!]
\label{table:car}
\begin{center}   
\begin{tabular}{||c | c c||} 
\hline
\hline
{\bfseries Characteristics} & &\\
\hline
\hline
{\bfseries Number of arrays} & 2  & 1 \\
\hline
{\bfseries Wavelength} [mm] & 1.15 & 2 \\
{\bfseries Pixel size} [mm] & 2  &   2.3 \\
{\bfseries Hilbert pattern} & yes & yes \\
{\bfseries Angular size} [$F_{lambda}$] & 1 & 0.7 \\
{\bfseries Total pixel number} & 1140 & 1020 \\
\hline 
\hline
{\bfseries Requirements (Goals)} & &\\
\hline
\hline
\hline
{\bfseries Valid Pixels} [\%] & 50 (90) & 50 (90) \\
{\bfseries F.O.V.} [arcmin] & 5 (6.5)  &   5 (6.5) \\
{\bfseries FWHM} [arcmin] & 12 (10) & 18 (16) \\
{\bfseries NEFD} [mJy$ \cdot $ s $^{0.5}$/beam] & 15 (30) & 10 (20) \\
\hline 
\hline
\end{tabular}
\end{center}
\caption{Principal characteristics (currently installed in NIKA~2 instrument), requirements and goals of the NIKA~2 instrument. } 
\end{table} 

\section{INTRODUCTION}
\label{sec:intro}  

New challenges in millimetre wave astronomy require high sensitivity and high resolution instruments. To achieve such a goal, the development of a new generation of array of detectors is needed, because current detectors such as high impedance bolometers are already photon noise limited both for space and for ground-based observations. One of the proposed technological solution is the use of Lumped Element Kinetic Inductance Detectors (LEKID) that allow for a large multiplexing factor frequency domain readout and an accessible manufacturing. This technological solution has been selected for the NIKA project that aims at constructing a dual-band millimetre camera for observation at the 30~m IRAM telescope (Pico Veleta, Spain).
The final goal of the project is to install the most powerful ever continuum instrument with new features never available before at the IRAM 30~m telescope (large FOV, multi-band, polarisation capabilities, high resolution).

\begin{figure}[h]
\begin{center}
\includegraphics[angle=0,width=0.75\textwidth]{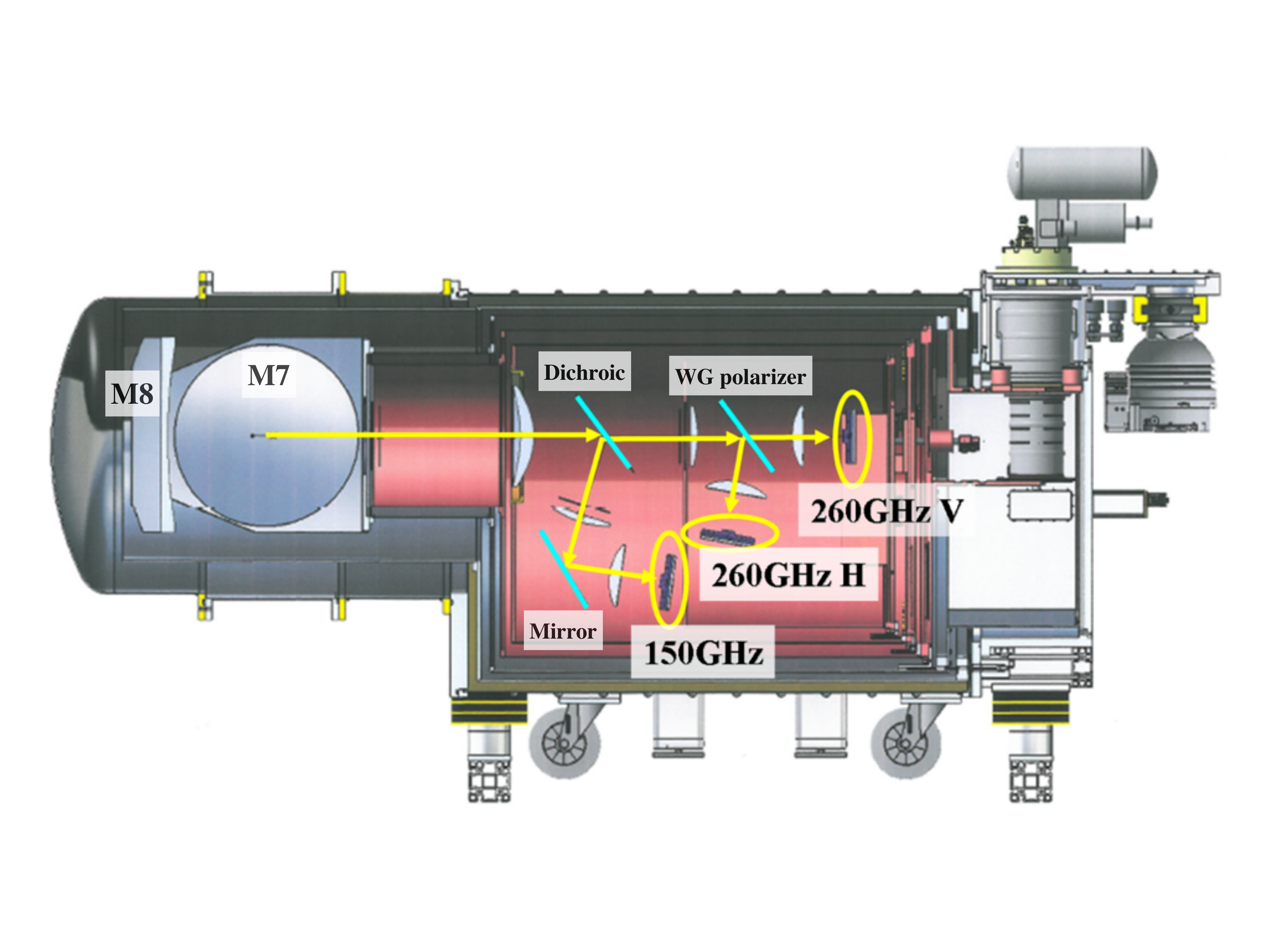}
\includegraphics[angle=0,width=0.4\textwidth]{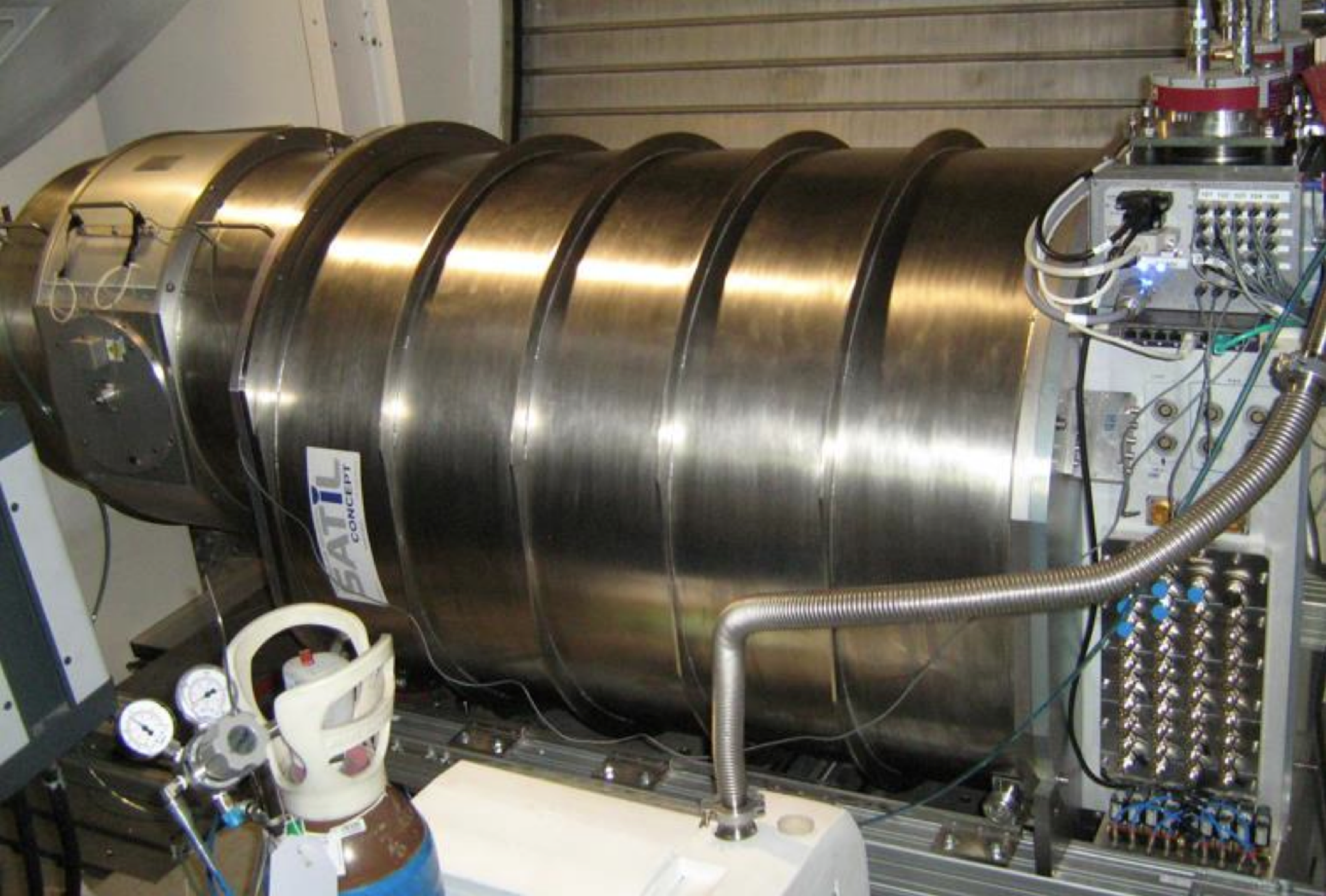}
\caption{Top panel: NIKA~2 schematic view of the cryostat with the cold optics. Bottom left panel: picture of the NIKA~2 cryostat installed at the IRAM 30~m receiver cabin.
\label{fig:fig1}}
\end{center}
\end{figure}

\begin{figure}[h!]
\begin{center}
\includegraphics[angle=0,width=0.7\textwidth]{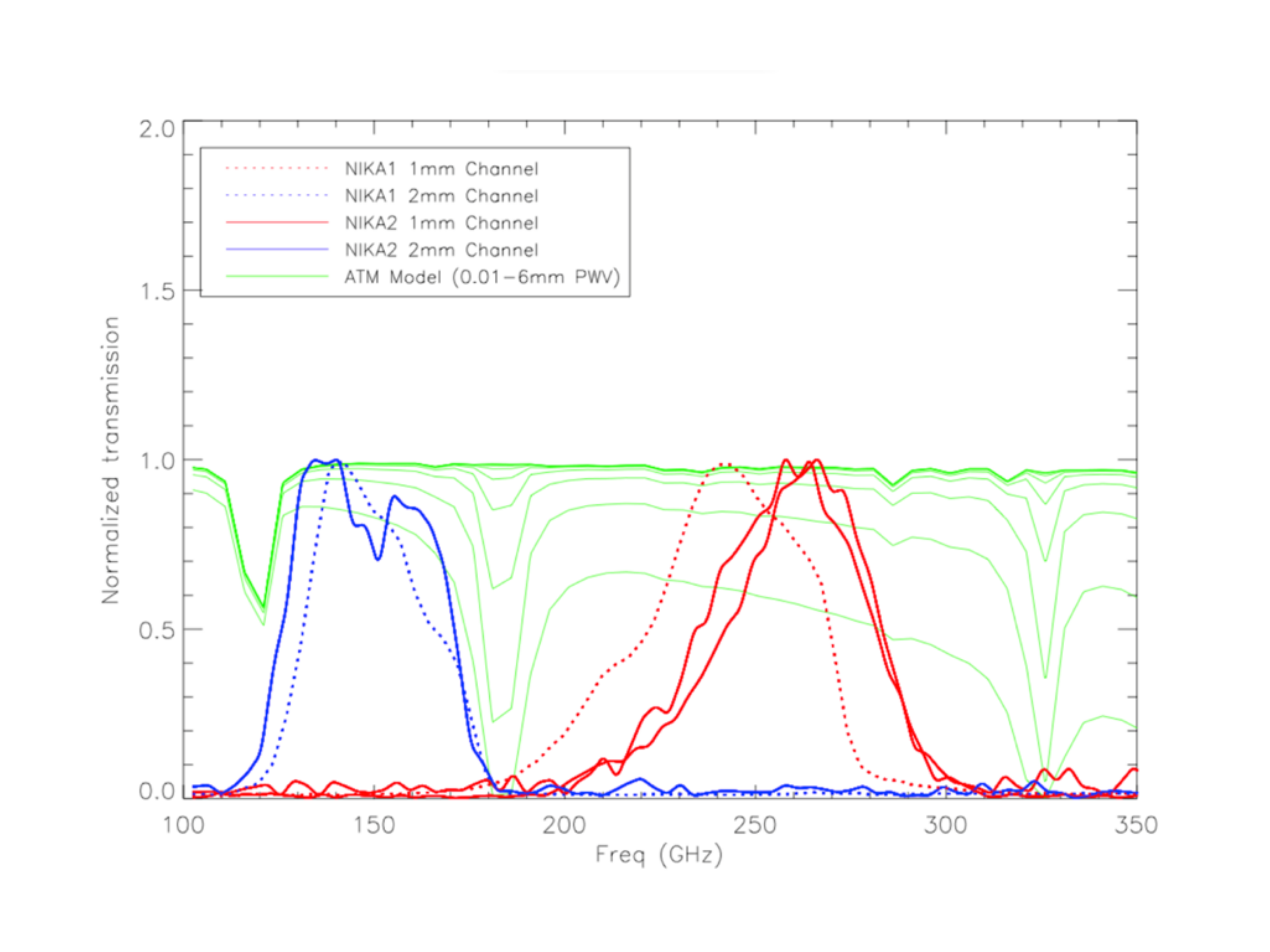}
\caption{ NIKA~2 bandpasses averaged over all valid pixels for 1.15~mm channel (red) and 2~mm channel (blue) compared to the ones of the NIKA (dotted lines). The atmospheric transmission modelled with ATM\cite{2001IEEE....49.1683C} for different calculated water vapor contents is presented in green.
\label{fig:fig11}}
\end{center}
\end{figure}

The camera has been permanently installed at the IRAM 30~m telescope in October 2015, and is expected to be available for the general community in 2017. The targets of the NIKA~2 camera are to perform simultaneous observations in two millimetre bands (1.15~mm and 2.0~mm) of sub-mJy point sources as well as to map extended continuum emission up to about 6.5~arcmin scale and beyond  with diffraction limited resolution and background limited performances. These specifications will make NIKA~2 an revolutionary instrument to map out Sunayaev-Zel'dovich signature in clusters out to distances of z > 2, effectively search for high redshift galaxies out to the range of the re-ionization epoch and finally map dust in nearby galaxies and galactic star forming regions. . In addition, the NIKA~2 instrument hosts polarisation capability in the 1.15 mm arrays. The adopted solution is the use of a rotating warm Half-Wave-Plate to modulate the astrophysical polarised signal and a polariser mounted at the 100~mK stage to analyse the linear polarisations on the two 1.15~mm arrays. 


\section{The NIKA~2 instrument and the integration at the 30~m IRAM telescope}\label{sec:nika}

The NIKA~2 instrument is essentially composed by four main sub-systems. We describe with some detail each of these sub-systems in the following: 


\begin{itemize}

\item {\bfseries Optics:}
One of the most important improvements between the NIKA and the NIKA~2 instrument is the change of the Field-of-View from about 1.9 to 6.5~arcmin. This permits to increase the mapping speed by a factor of ten if all the other sub-system performance are kept constant. To reach this goal the mirrors of the receiver cabin (M3, M4, M5 and M6 being M1 and M2 the primary 30~m mirror and M2 the sub-reflector) have been modified in shape and in size in order to avoid the vignetting of such a big field of view. Inside the imager, the optical coupling between the telescope plus receiver cabin mirrors and the detectors is made by warm aluminium mirrors (M7 and M8 cooled at 80~K) and cold refractive optics (high density polyethylene - HDPE lenses placed from the 1~K stage to the 100~mK stage).

The rejection of out-of-band emission from the sky and the telescope is achieved by using a series of low-pass metal mesh filters, placed at different cryogenic stages in order to
minimize the thermal loading on the detectors. The band splitting between 2 and 1.15~mm channels is achieved by a dichroic installed at the 100~mK stage.  In addition to dual-polarisation observations, NIKA~2 has polarisation capabilities. Optically, they consist of a multi-mesh hot-pressed Half-Wave-Plate (HWP)\cite{pisano2016} mounted at few centimetres from the NIKA~2 cryostat window in a mechanical modulator performing the rotation thanks to a step-by-step motor. Since the LEKID design used in the NIKA~2 instrument are sensitive to both polarisations, the beam is further split by a wire grid polariser at 100 mK cryogenic stage in order to analyse the modulated linear polarisation into two arrays observing at 1.15~mm  without any loss of signal. We present in Fig.\ref{fig:fig1} a schematic representation of the NIKA~2. 

Spectral characterisation of the NIKA bandpass was performed in laboratory using a Martin-Puplett interferometer allowing recovery of the spectral performance of each pixel of the two NIKA~2 channels with uncertainties of a few percent. In Fig~\ref{fig:fig11}, we present the NIKA bandpasses, together with the Atmospheric transmission at microwaves (ATM) model calculated for different water vapour contents\cite{2001IEEE....49.1683C}.

\begin{figure}[h!]
\begin{center}
\includegraphics[angle=0,width=0.45\textwidth]{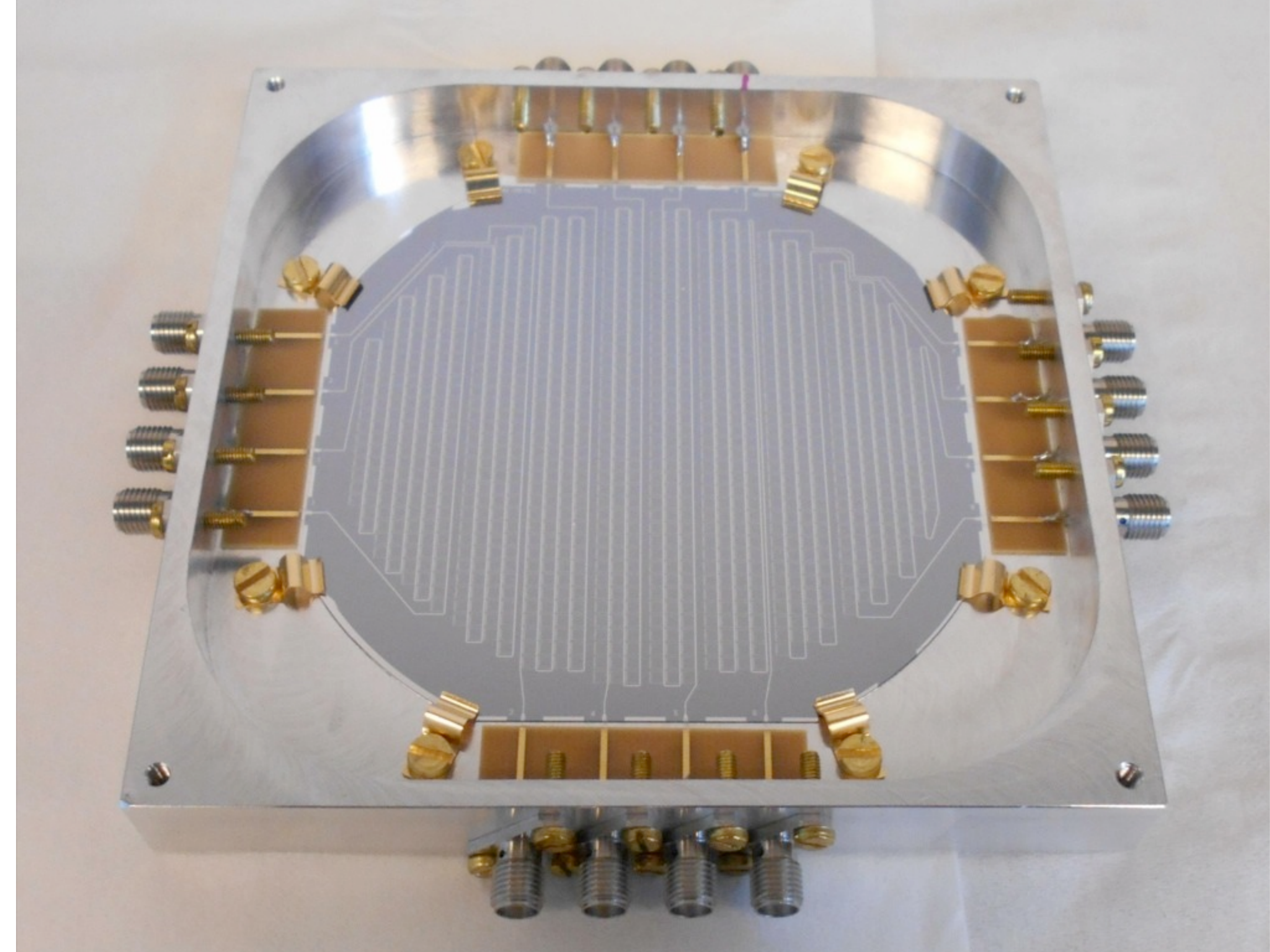}
\includegraphics[angle=0,width=0.45\textwidth]{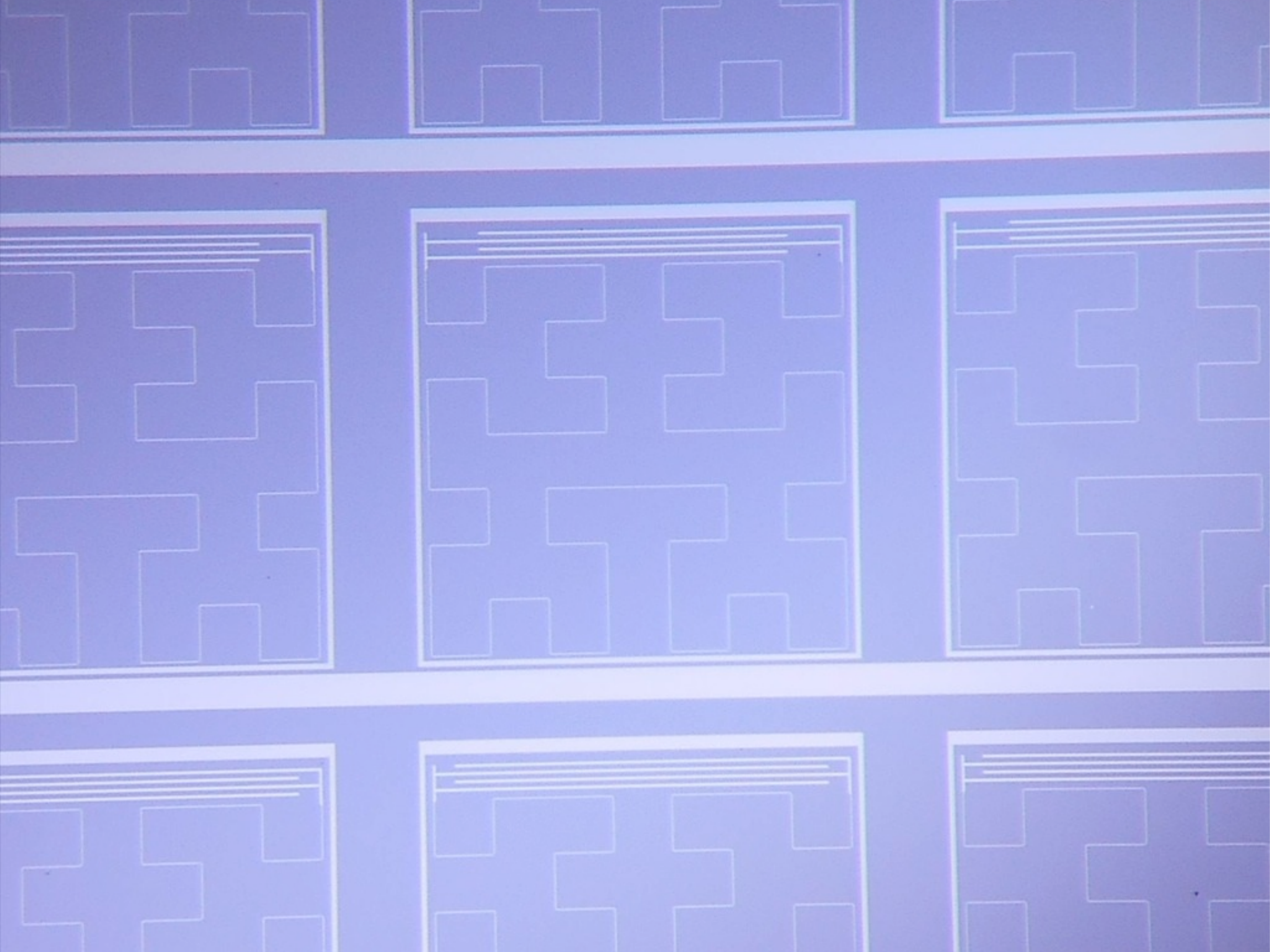}
\caption{Left: picture of the NIKA~2 1.15~mm module array. Right: zoom on 1.15~mm channel array. The resonators are composed of a long inductive meander line based on Hilbert-shape and a capacitive element that is also used for tuning the resonant frequency. 
\label{fig:fig2}}
\end{center}
\end{figure}

\item {\bfseries Cryostat:}
Cooling the three KID arrays at a nominal temperature of 150~mK is the major requirement that drove the architecture of the NIKA~2 instrument. The cryostat has been designed at Institute Neel and it consists of about 3000 mechanical pieces for a total weight of 1.3~tons. The nominal working temperature is achieved by a 4~K cryocooler and a closed-cycle $^3He$-$^4He$ dilution fridge. First, the cryostat is pre-cooled to about 4-5~K by two Pulse Tubes with cooling capacity of 1.35~Watt each and working in parallel. The pulse tube pipes are 60 meters long and run through the telescope's cable spiral in order to connect the heads in the receiver cabin (rotating in azimuth) and the compressors located in the pedestal.
Then, the final working temperature is achieved by the dilution refrigerator. 

The cool-down process is completely remote controlled and it does not use any cryogenics liquids. The whole process lasts about 5 days with four full days of pre-cooling and about one day of dilution cool-down.

\begin{table}[b!]
\label{table:reu}
\begin{center}       
\begin{tabular}{||c | c c||} 
\hline
\hline
{\bfseries Wavelength} [mm] & 2 & 1.15 \\
\hline
\hline
{\bfseries Average KID per feed-line}   [\#] & 255   &     142.5 \\
{\bfseries Board count}  [\#] & 4   &     16  \\
{\bfseries Power consumption}  [W] & 370  &   1220 \\
{\bfseries Tone tuning resolution} [Hz] & 953 &  953 \\
{\bfseries Frequency range}  [GHz] &1.3-1.8 &        1.9-2.4 \\
\hline 
\hline
\end{tabular}
\end{center}
\caption{Main characteristics of the NIKA~2 readout.} 
\end{table}

\item {\bfseries Detectors:}

NIKA~2 detectors are Lumped Element Kinetic Inductance Detectors (LEKIDs). Each pixel is a resonator composed of an interdigitated capacitor and a meander inductor (acting as an absorber) designed with an Hilbert pattern to absorb both polarisations\cite{Roesch2012,Roesch2}. 
Each array is fabricated on a single 4 inches high resistivity silicon wafer (thickness equal to 250 and 300~$\mu m$ for 1.15~mm and 2~mm respectively). Pixels are wet etched from a thin aluminium film (18~nm) deposited by e-beam evaporation. The use of thin aluminium film has several advantages: a better match with free space impedance of the incoming photons and a high kinetic inductance. This kind of film has been largely adopted for NIKA and in general for all the applications in our laboratory\cite{goupyLTD,catalano_3mm,catalano_space}. 
In order to exploit all the angular resolution of the IRAM 30~m telescope, the pixel sizes have been set to 2~mm for the 1.15~mm arrays and 2.3~mm for 2~mm array. This corresponds to F$\lambda$ equal to 1 and 0.7 for 1.15~mm and 2~mm arrays, respectively. 
The geometrical coupling between pixels and the corresponding feed-line is made with a microstrip readout line. This solution represents an improvement with respect to the solution chosen for the NIKA (Coplanar Waveguide - CPW) giving more homogeneous and predictable performances. Since the back-side of the wafer is metallized in this configuration, the consequence of this geometry is that the detectors must be front-illuminated. More details about design and fabrication of NIKA detectors can be found in \cite{goupyLTD, Roesch2012 ,calvoLTD}.

\begin{figure}[h!]
\begin{center}
\includegraphics[angle=0,width=1\textwidth]{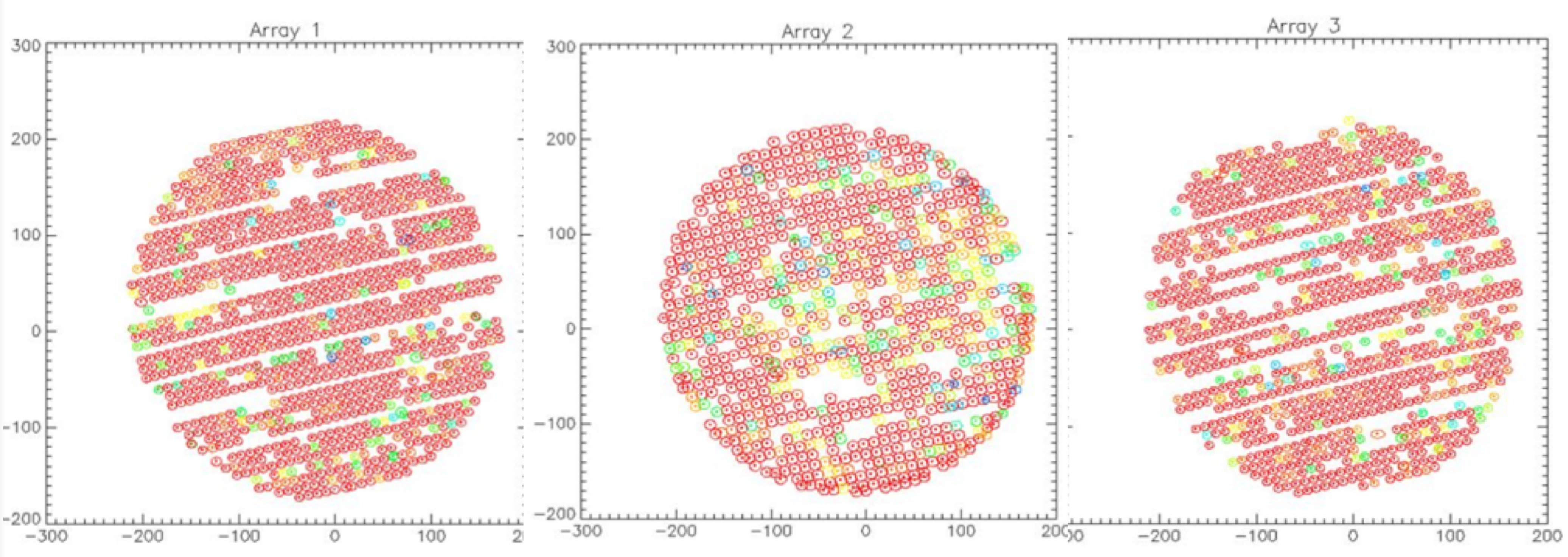}
\caption{Field Of View (FOV) reconstruction using point source Uranus.They were obtained combining several observations performed in March 2016. The color shows the stability along the run (the redder the better). X and Y axis are arc-seconds.
\label{fig:fig3}}
\end{center}
\end{figure}

\item {\bfseries Readout electronics:}
The readout comprises coaxial cables connected from 300~K to the base temperature, 20 low-noise cryogenics amplifiers (LNA) installed in the 4~K stage, and warm electronics. The latter is constituted by readout boards named New Iram Kid ELectronic in Advanced Mezzanine Card format (NIKEL\_AMC), central, clocking and synchronisation boards (CCSB) mounted on the MicroTCA Carrier Hub (MCH) and one 600\,W power supply. These boards are distributed in three micro-Telecommunication Computing Architecture (MTCA) crates. In order to readout 3300 pixels, NIKA~2 is equipped with 3 crates hosting 20 boards (8X2 for arrays at 1.15~mm and 4x1 for 2~mm array). 
\\
A NIKEL\_AMC board works with 6 separate Field Programmable Gate Arrays (FPGAs) each coupled to a Digital to Analog Converter; they can generate a comb of frequencies each over a 500~MHz bandwidth each set to the resonant frequency of a LEKID in a feedline. The comb is then up-converted by mixing with a local oscillator carrier at the appropriate frequency. The output line is down-converted to the base band and is acquired by an Analog to Digital Converter (ADC), and then compared to a copy of the input tones kept as a reference in order to extract the in-phase (I) and in-quadrature (Q) signal. 
\\
For a full description of the readout electronics please see for example\cite{Bourrion2012, bourrion2016}. The main characteristics of the readout are presented in Table~\ref{table:reu}.

\end{itemize}

\subsection {The beginning of the commissioning phase}
As shown in Tab~1, the installation of the NIKA~2 at the IRAM 30~m telescope was performed in the beginning of October 2015 and lasted few days. During the night between the 6th and 7th of October 2015, the instrument observed its first light on Uranus. Commissioning has started at that time is expected to last until fall of 2016. The NIKA~2 collaboration is responsible for the commissioning and the documentation of the instrument. The complete documentation and transfer of required operation and maintenance skills to the IRAM staff are part of the commissioning phase goals.


\section{Calibration process}\label{sec:cal}
The variety of NIKA~2 scientific targets requires a dedicated readout technique, an accurate photometric calibration and a specific data reduction pipeline. The NIKA~2 calibration process follows closely the one developed for the NIKA which has been extensively described in \cite{NIKA2014, NIKA_SZ}. {\bfseries The results presented in this work are preliminary depending on the fact that the data analysis of the NIKA~2 commissioning phase are still in progress}. 

We list in the following the critical points to be controlled:

\begin{figure}[h!]
\begin{center}
\includegraphics[angle=0,width=0.45\textwidth]{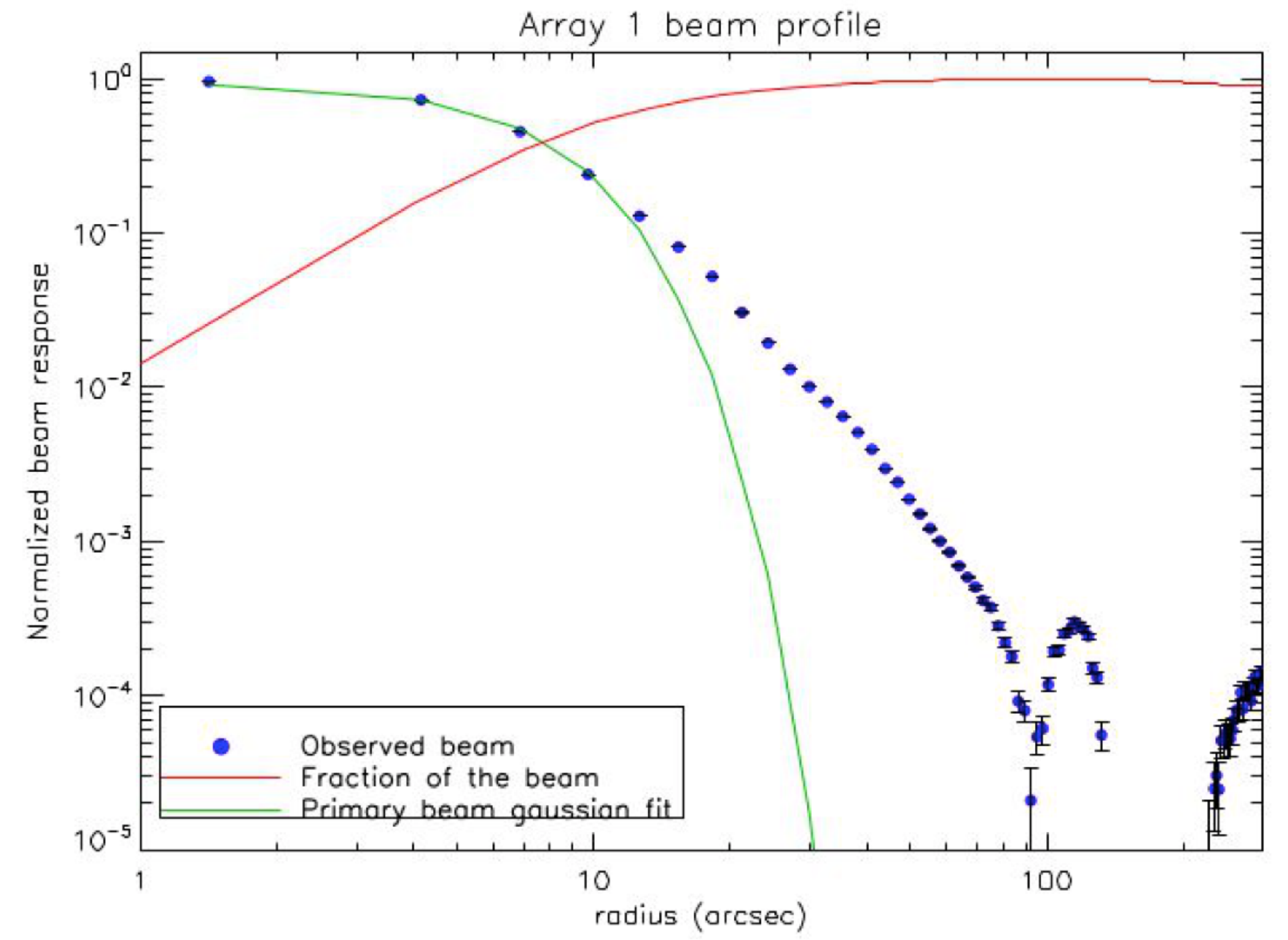}
\includegraphics[angle=0,width=0.45\textwidth]{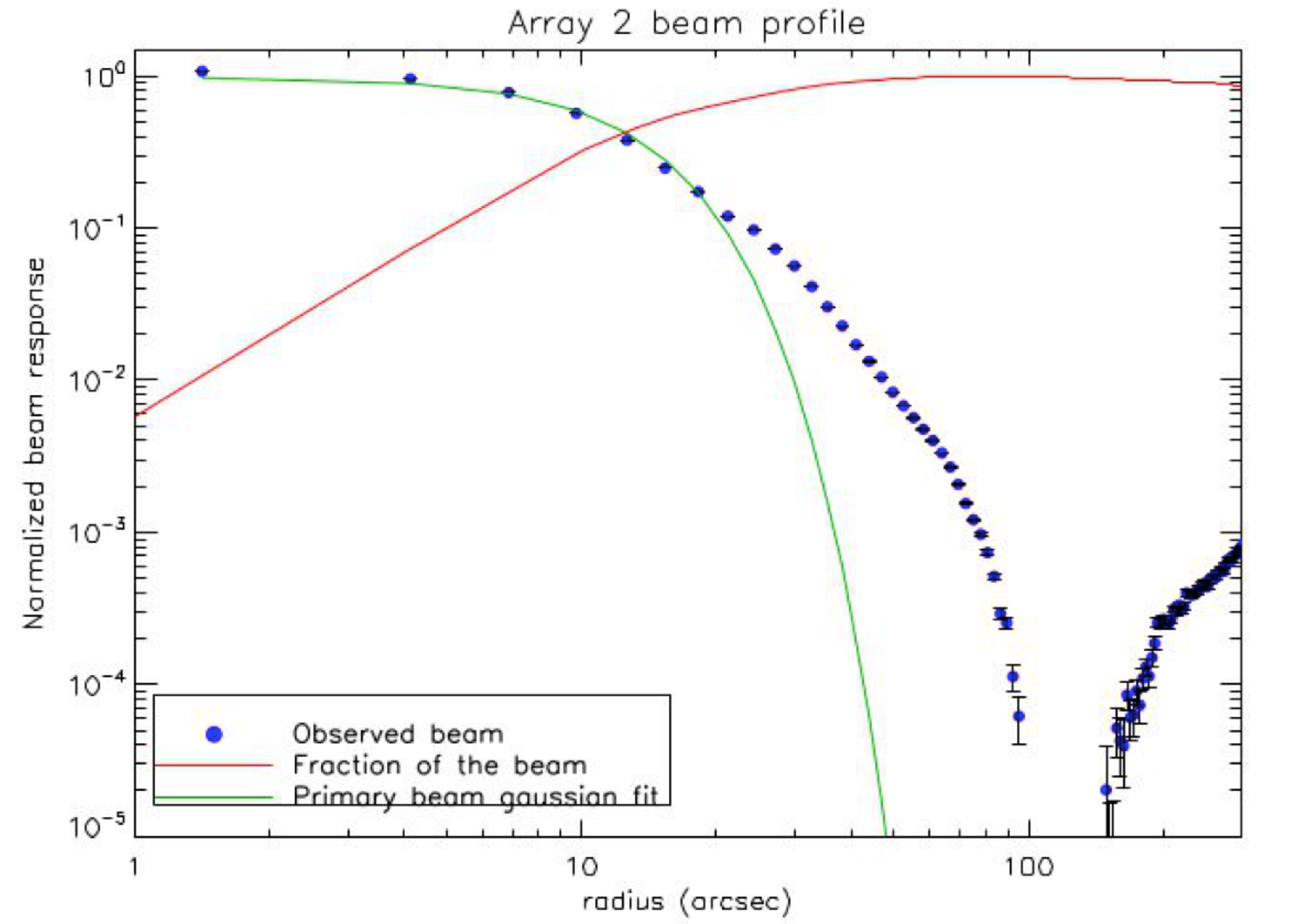}\\
\includegraphics[angle=0,width=0.45\textwidth]{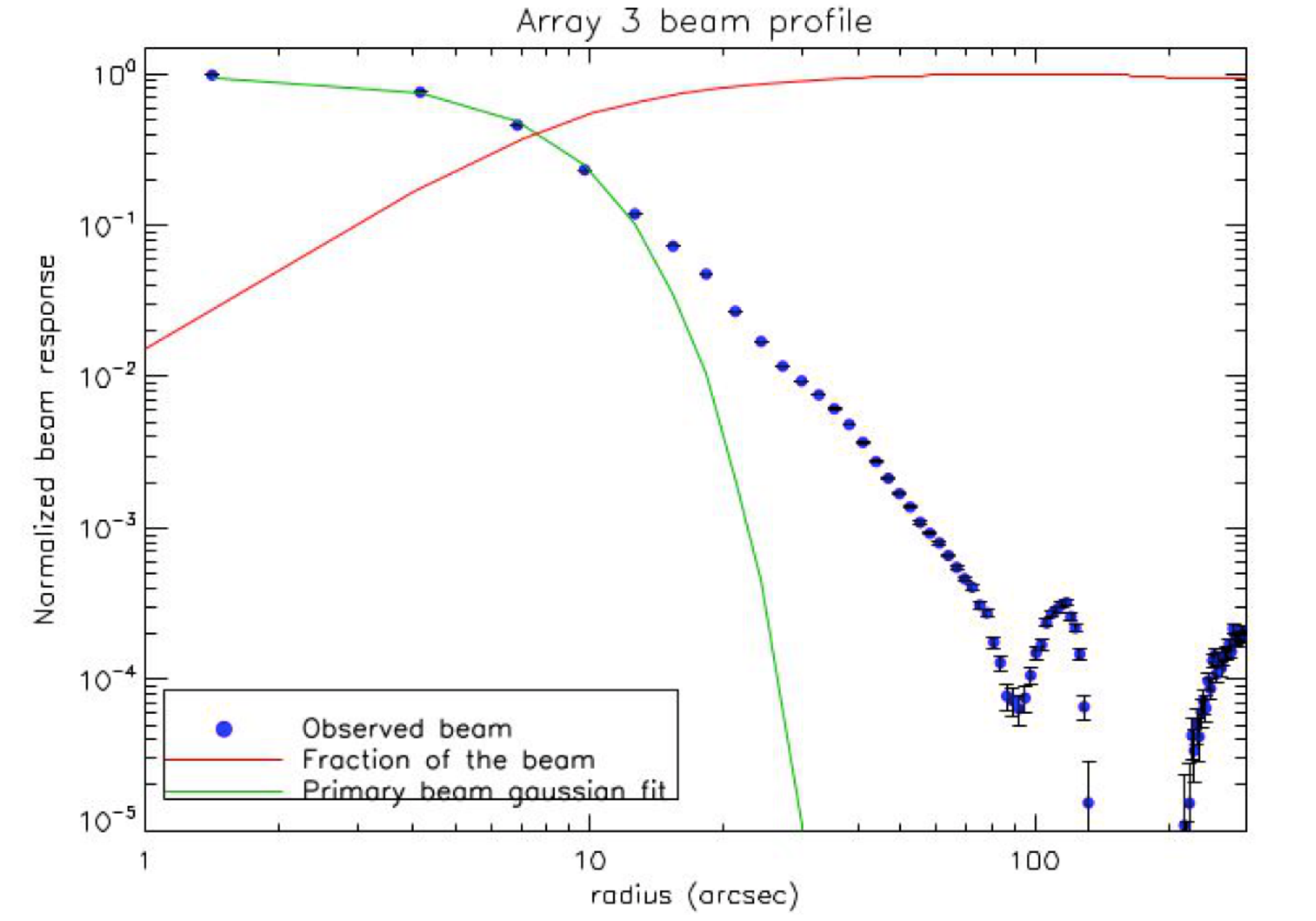}
\caption{Beam profiles obtained from point source observations. We compute a map profile and fit a 2D Gaussian to it. The blue points are the measured data from the point source maps, the green line is the Gaussian best-fit and the red ones correspond to  the integrated beam as a function of the distance from the beam centre. The main beam is to first order consistent with the typical beam pattern of the 30~m telescope as well as with the performance obtained by the NIKA.
\label{fig:fig6}}
\end{center}
\end{figure}

\begin{itemize}

\item {\bfseries Readout optimisation technique:} One of the most difficult challenges in operating with KIDs, is to convert the observed in phase (I(t)) and in quadrature (Q(t)) signal to absorbed optical power. If the sky emission fluctuates during a scan, the resonance circle changes and therefore the responsivity of the detector changes. To improve the photometric reproducibility we developed a system to control the change of the signal by modulating the frequency on the local oscillator (by few kHz) synchronously to the acquisition cycle in order to generate two tones, one juste below and the other juste above the resonant frequency of the detector. An average of 40 points (or 20 for polarisation observations) are then recorded by the FPGA acquiring data at the rate of about 23~Hz (or 46~Hz for polarisation observations). In this method we estimate the variation of the resonant frequency of the detectors $\Delta f_0(t)$, by projecting $(\Delta I(t), \Delta Q(t))$ along the gradient. A detailed description of this method is given in \cite{Calvo2013, NIKA2014}.
\\
Another tool developed to optimize the detectors working point before each astrophysical observation, is the tuning procedure:  
for a ground-based instrument, the variations of the optical background due to the atmosphere shifts the resonant frequency of the detectors by a substantial amount (in some cases larger than the resonances themselves). This effect is constantly monitored and, if the resonant frequency crosses the defined threshold, it can be balanced by changing the excitation tones, in order to always match the resonant frequency of each pixel and to thus probe each detector near to its ideal working point. The tuning procedure is based on the measurement of the angle $\varphi$ between the vectors ($I$, $Q$) and ($dI/df_{LO}$, $dQ/df_{LO}$). Using this angle we are able, with a single data point, to retune the detectors. This represents a crucial advantage in terms of observing time, especially in the case of medium and poor weather conditions. A detailed description of this method is given in\cite{NIKA2014}.

\begin{figure}[h!]
\begin{center}
\includegraphics[angle=0,width=0.55\textwidth]{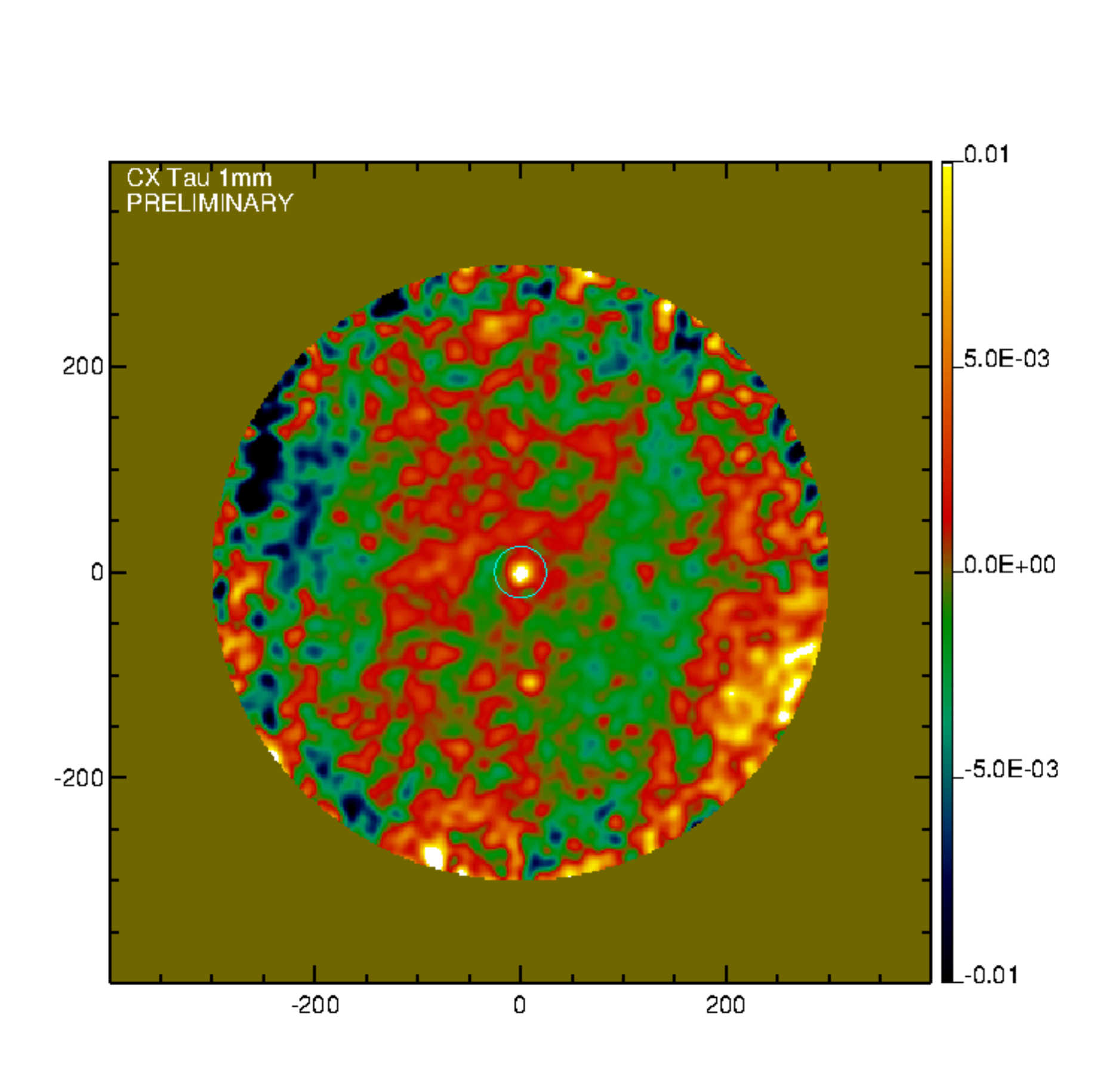}
\caption{Preliminary flux map of CXTau at 1.15~mm observed in about 1.5~h. The 10~mJy source is well detected, in line with the instrument specifications on sensitivity.
\label{fig:fig5}}
\end{center}
\end{figure}

\item {\bfseries Photometric calibration:}  The photometric calibration in NIKA~2 pipeline consists in a series of procedures allowing us to characterise the accuracy of the instrument photometry and to define the impact of systematic errors in the final sky maps. Uncertainties come from both external (calibration factors deduced from primary calibrator sources) and internal instrument uncertainties (knowledge of pointing directions for each pixel, spectral response uncertainty, secondary beam fraction and opacity correction). 
In Fig~\ref{fig:fig3} we show the reconstructed focal plane geometry for the three NIKA2 arrays. The FOV is confirmed to be 6.5~arcmin for the three arrays. Each circle corresponds to one pixel, and the different colours are associated to the stability over the different observations. Pixels having a too large noise or too elliptic beams are not considered for the analysis. The level of valid pixels is between  80 and 90~\% for the 1.15~mm arrays and between 70 and 90~\% for the 2~mm array with a measured FWHM of about 12'' and 18'', respectively. A more detailed characterisation of the pixel response is being performed and it is one of the major goals of the remaining NIKA2 commissioning analysis. To date we estimate that the overall calibration uncertainty for point sources on the final data at the map level is around between 8 and 15~\% for 1.15~mm channels and 7 and 10~\% for 2~mm channel. This is based on the observed root mean square flux variation on a calibration source. A more in depth analysis is being carried out in order to stabilise and homogenise the results. 

\item {\bfseries Data Reduction : } 
As well as producing calibrated data, the reduction pipeline stores, filters and processes data onto sky maps. Roughly speaking, raw data are recorded, unstable detectors are flagged out depending on their statistical noise properties. Data are calibrated (accounting for opacity). A dedicated Fourier space filtering is used to remove frequency lines produced by the vibrations of the pulse tube. Finally atmospheric and electronic noises are decorrelated and the time ordered data projected onto maps.Each of these steps is made complex by the number of detectors and the readout architecture that introduces different sources of correlated noise. These features were expected from the instrument design and our data analysis pipeline is designed to cope with them. 

As an illustration, we show two preliminary maps obtained during the first few weeks of commissioning that illustrates the ability of NIKA2 to map faint point sources and extended structures. In Fig.~\ref{fig:fig5} we present faint T-Tauri star CSTau and its disk (about 10~mJy at 1.15~mm), to illustrate NIKA~2's potential sensitivity. 
In Fig~\ref{fig:dr21oh} we present a first light image obtained on DR21OH with a integration of few minutes. DR21OH is a well-known star forming region inside the Cygnus X molecular cloud complex.

\begin{figure}[h!]
\begin{center}
\includegraphics[angle=0,width=0.85\textwidth]{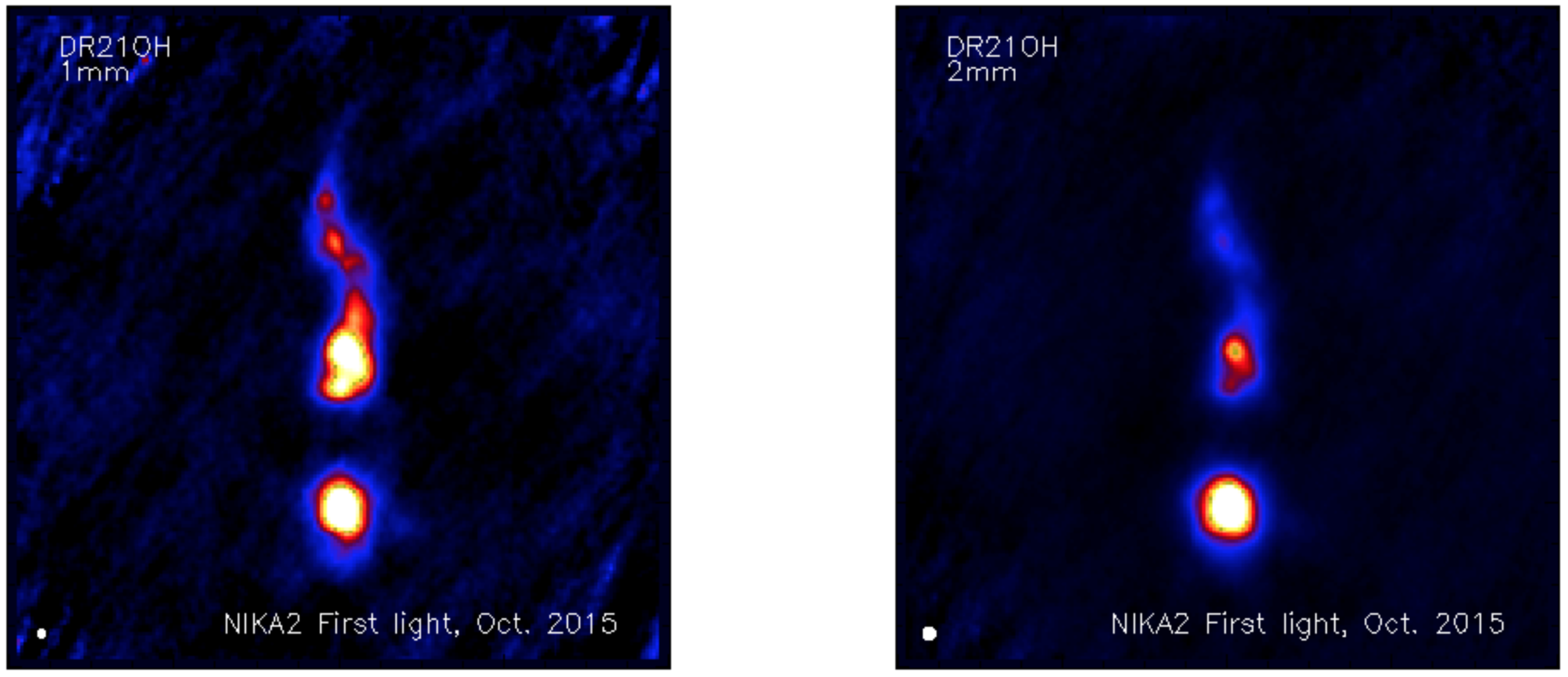}
\caption{Preliminary map of DR21OH observed at 1.15 and 2~mm. Maps are 13 arcmin large, the angular resolution in each band is represented as a white disk at the bottom of the images.
\label{fig:dr21oh}}
\end{center}
\end{figure}

\item {\bfseries Polarisation Data Reduction : } 
As discussed above NIKA~2 has polarisation capabilities for the 1.15~mm channel and permits the measurement of the three Stokes parameters representing linear polarisation, $I$, $Q$ and $U$.
In addition to the intensity procedures presented above, we have developed specific polarisation ones, as the reconstruction of the position of the HWP, the subtraction of the parasitic HWP synchronous signal, polarised map making and intensity-to-polarisation leakage correction. 
A description of the polarised data reduction pipeline is presented in \cite{ritacco}

\end{itemize}



\section{NIKA~2 commissioning: observing campaigns 2015-2016}\label{sec:run12-13-14}

In this section we give a panorama of the different observing campaigns performed from October 2015 to March 2016 together with the planned future tests and instrument upgrades.  

\begin{itemize}

\item {\bf 1$^{st}$ test run:} from 2015-10-29 to 2015-11-10. Installation and cool-down at the 30~m telescope. This first run started with a reduced number of readout electronic boards and therefore with a reduced number of pixels on the sky (8 feed-lines over the 20), but sufficiently complete to start the characterisation of the instrument. The main outcome of this run was the integration of all sub-systems to the telescope. In addition, we were able to accumulate some statistics on beam maps and on noise equivalent flux density (NEFD). 

\item {\bf 2$^{nd}$ test run:} from 2015-11-24 to 2015-12-02. Continuation of tests started during the first run with more readout boards (13 feed-lines covered). Many observations of beam maps and NEFD to increase statistics and confirm the first run results with a larger coverage of the focal plane.

\item {\bf 3$^{rd}$ test run:} from 2016-01-12 to 2016-02-01. Implementation of new NIKEL boards; the 20 feed-lines of NIKA2 have been plugged simultaneously for the first time. With the 3 arrays fully equipped, we continue to perform the tests started with runs 1 and 2. First observations in polarisation mode show that the polarisation facilities works good. Accurate polarisation data analysis is postponed to next months.  

\item {\bf 4$^{th}$ test run:} from 2016-03-01 to 2016-03-15. Deeper beam patterns, array 1 and 3 performances versus electronic boards, NEFD, mapping strategies, statistics accumulation, and polarisation observations of quasars and polarised extended sources.
The beam morphology has been characterised showing a side lobe compatible with the NIKA side lobes (see Fig~\ref{fig:fig6}). 
The distortion of the focal plane image has been characterised showing inhomogeneous distortions over the three different arrays. The outcomes of photometric calibration accuracy and sensitivity are not consolidated yet, but it seems to be in line with requirements. 

\item {\bf Future test runs and upgrades:} first commissioning observing campaigns have permitted to the collaboration to test the capabilities of NIKA 2 instrument and to confirm that the instrument is in line with requirements. As expected, more work is necessary to consolidate all the results. The collaboration is working hard to this task. Few systematic effects have been isolated and deeply studied. In order to minimise these effects, we planned to upgrade the instrument in September 2016 by replacing or upgrading the instrument with few sub-systems: 

1) Replace the 2~mm array to improve cosmetics.

2) Adding dark pixels in small arrays connected in series to each "illuminated" feed-line to monitor more efficiently the correlated noise.

3) Ugrade of few readout electronics boards with the last NIKEL version to improve the noise homogeneity over the arrays. 

4) Replace all the lenses and the cryostat window adding geometrical anti-reflection coating. This will improve the optical efficiency. 

5) Replace the dichroic to minimize the distortion of the focal place image. 

\end{itemize}

\section{Conclusion} 
NIKA2 has been successfully installed at the IRAM 30~m telescope in october 2015 as planned. The commissioning phase is still in progress and it is showing very promising results in prospect of the opening of the instrument to the community in 2017 as scheduled. The first four test runs have proved that all the sub-systems work nominally and that specifications should be met. Few systematic effects have been isolated (low level beams distortion, stability, data analysis software), which we expect to minimise by an upgrade of the instrument planned for September 2016. 



\section*{\small ACKNOWLEDGMENTS}       
\footnotesize
We would like to thank the IRAM staff for their support during the campaigns. 
The NIKA~2 dilution cryostat has been designed and built at the Institut N\'eel. 
In particular, we acknowledge the crucial contribution of the Cryogenics Group, and 
in particular Gregory Garde, Henri Rodenas, Jean Paul Leggeri, Philippe Camus. 
This work has been partially funded by the Foundation Nanoscience Grenoble, the LabEx FOCUS ANR-11-LABX-0013 and 
the ANR under the contracts "MKIDS", "NIKA" and ANR-15-CE31-0017. 
This work has benefited from the support of the European Research Council Advanced Grant ORISTARS 
under the European Union's Seventh Framework Programme (Grant Agreement no. 291294).
We acknowledge fundings from the ENIGMASS French LabEx (R. A. and F. R.), 
the CNES post-doctoral fellowship program (R. A.),  the CNES doctoral fellowship program (A. R.) and  the FOCUS French LabEx doctoral fellowship program (A. R.).
G.L., A.B. et H.A. acknowledge financial support from "Programme National de Cosmologie and Galaxies" (PNCG) of CNRS/INSU, France, and from OCEVU Labex(ANR-11-LABX-0060)
\small
\bibliography{cataSPIE}   
\bibliographystyle{spiebib}   

\end{document}